\documentclass[a4paper]{jpconf}
\usepackage{graphicx}
\usepackage{bm}
\usepackage{subfigure}
\usepackage{amsmath,amssymb}
\usepackage{xcolor}

\begin{document}
\title{Theory of Defect-Induced Kondo Effect in Graphene:
\\Effect of Zeeman Field}

\author{Taro Kanao, Hiroyasu Matsuura and Masao Ogata}

\address{Department of Physics, the University of Tokyo, Bunkyo, Tokyo 113-0033, Japan}

\ead{kanao@hosi.phys.s.u-tokyo.ac.jp}

\begin{abstract}
The effect of the Zeeman field on the defect-induced Kondo effect in graphene is investigated. 
The effective model of the Kondo effect is derived, and is analyzed on the basis of the numerical renormalization group method. 
It is found that, under the Zeeman field, at the border of the spin polarized state and the Kondo-Yosida singlet state, there is an unusual fixed point where the entropy of the defect is $k_B\ln2$ and the expectation value of $S^2_z$ is $1/8$. 
\end{abstract}

\section{Introduction}
The effects of defects on the properties of graphene are of scientific and practical importance~\cite{Terrones2012}. 
For studying the effects of defects experimentally, ion irradiation has been used, which introduces point defects in graphene. 
Recently, in such an ion-irradiated graphene, the Kondo effect is observed ~\cite{Chen2011}. 
Usually, the Kondo effect indicates the presence of localized moments in the sample. 
Therefore the above result suggests that a point defect in graphene causes a localized moment. 
The magnetic behavior of the defects in graphene is also observed in the recent magnetization measurement of ion-irradiated graphene~\cite{Nair2012}. 

As a possible origin of this Kondo effect, we proposed a localized moment on a $sp^2$ orbital around the defect ~\cite{Kanao2012}. 
There are three $sp^2$ orbitals around a point defect. 
With Jahn-Teller distortion, two of them form a covalent bond and remaining one behaves as a localized moment with spin-half. 
Furthermore, with this lattice distortion, there is a hybridization between the localized $sp^2$ orbital and conduction $\pi$ electrons, and the Kondo effect can occur.  
On this scenario, we introduced an effective model and analyzed it by the numerical renormalization group (NRG) method~\cite{Wilson1975, Bulla2008}. 
As a result, we showed that the experimentally observed gate voltage dependence of Kondo temperature can be understood by assuming that the hybridization between the $sp^2$ and $\pi$ orbitals is strong enough~\cite{Kanao2012}.  

In the above experiment, the negative magnetoresistance (MR), which is the evidence of the Kondo effect, was also observed. 
Although the characteristic magnetic field for this negative MR is found to be smaller by one order than the usual Kondo-effect case, its origin has not been understood. 
So far, the studies on the effect of magnetic field on the Kondo effect in graphene are limited, and there are few studies on the effect of Zeeman field~\cite{Cornaglia2009}. 

Motivated by this negative MR, in this paper, we construct an effective model of graphene with a point defect and study the magnetic field effect on the thermodynamics of the model on the basis of the NRG method. 
Here, as a first step, we consider only the Zeeman field and neglect the orbital magnetic effect on the conduction electrons. 
 
\section{Model Hamiltonian with the Zeeman field}
The effective Hamiltonian of the graphene with a point defect has been discussed in ref.~\cite{Kanao2012}, and the one-dimensional energy representation for the conduction electrons~\cite{Bulla2008, Jones1987, Sakai1992} which is necessary to apply the NRG method has been derived. 
In this representation, the effective Hamiltonian is written as 
\begin{eqnarray}
	H&=&H_{\rm{gra}}+H_{\rm{def}}+H_{\rm{hyb}}, \label{eq_Hamiltonian}
\end{eqnarray}
where $H_{\rm{gra}}$, $H_{\rm{def}}$, and $H_{\rm{hyb}}$ are Hamiltonians for the conduction electrons, the defect $sp^2$ orbital, and the hybridization, respectively, with
\begin{eqnarray}
	H_{\rm{gra}}&=&\sum_{\sigma=\uparrow, \downarrow}\int^{D}_{-D}\mathrm{d}\varepsilon\left(\varepsilon-\mu\right)c^\dagger_{\varepsilon\sigma}c_{\varepsilon\sigma}, \label{eq_HamiltonianGraphene}\\
	H_{\rm{def}}&=&\sum_{\sigma}(\epsilon_{sp^2}-\mu)d^\dagger_{\sigma}d_{\sigma}+Ud^\dagger_{\uparrow}d_{\uparrow}d^\dagger_{\downarrow}d_{\downarrow}-\sum_\sigma h\sigma d^\dagger_\sigma d_\sigma,\label{eq_HamiltonianDefect}\\
	H_{\rm{hyb}}&=&\frac{V}{D}\sum_\sigma\int^{D}_{-D}\mathrm{d}\varepsilon|\varepsilon|^{1/2}\left(c^\dagger_{\varepsilon\sigma}d_\sigma+\rm{h.c.}\right).\label{eq_HamiltonianHybridization}
\end{eqnarray}
Here, $c_{\varepsilon\sigma}$ is an annihilation operator of the conduction electrons with energy $\varepsilon$ and spin $\sigma=\uparrow$, $\downarrow$, which satisfies the anti-commutation relation $\left\{c_{\varepsilon\sigma},c^\dagger_{\varepsilon'\sigma'}\right\}=\delta_{\sigma\sigma'}\delta(\varepsilon-\varepsilon')$~\cite{Kanao2012, Bulla2008, Jones1987, Sakai1992}. 
$\delta(x)$ is the Dirac delta function.
$D$ is a cut-off energy that is the same order as the band width of the conduction electrons. 
$d_\sigma$ is an annihilation operator of an electron on the defect $sp^2$ orbital with spin $\sigma=\uparrow$, $\downarrow$, which satisfies $\left\{d_\sigma,d^\dagger_{\sigma'}\right\}=\delta_{\sigma\sigma'}$.
$\mu$ is the chemical potential of electrons. 
$\epsilon_{sp^2}$, $U$, and $h$ are the energy level of the defect $sp^2$ orbital measured from that of $p_z$ orbital on carbon atom, Coulomb interaction, and Zeeman field on the defect $sp^2$ orbital, respectively. 
$V$ is the amplitude of the hybridization between conduction electrons and the $sp^2$ orbital. 

This model of eqs. (\ref{eq_Hamiltonian})-(\ref{eq_HamiltonianHybridization}) is known as the single-channel pseudogap Anderson model~\cite{Gonzalez-Buxton1998}, where the density of states (DOS) of the conduction electron is proportional to $|\varepsilon|$. 
The Kondo problem in this pseudogap system was studied in detail for the case with $\mu=0$ and $h=0$~\cite{Gonzalez-Buxton1998}. 
The main feature in this case is as follows. 
In the particle-hole symmetric ($2\epsilon_{ sp^2}+U=0$) case, no Kondo screening occurs, i.e. the localized moment is not screened by conduction electrons. 
On the other hand, in the particle-hole asymmetric ($2\epsilon_{ sp^2}+U\neq0$) case, the localized moment is screened when the coupling constant $V$ is larger than a critical value $V_c$. 
In this paper, we assume the particle-hole asymmetry and study the case with $h\neq0$ and both $\mu=0$ and $\mu\neq0$. 

\section{Numerical renormalization group method}
We apply the NRG method for the model (\ref{eq_Hamiltonian})-(\ref{eq_HamiltonianHybridization}), following the formalism of ref.~\cite{Gonzalez-Buxton1998} to investigate the electronic state at low temperatures. 
The NRG method can include the effects of the pseudogap DOS, a finite chemical potential, and Zeeman field. 
In the NRG method, the effective Hamiltonian is transformed into the one-dimensional form with a chain length $N$, 
\begin{eqnarray}
	H=\lim_{N\rightarrow\infty}\alpha\Lambda^{-(N-1)/2}H_N, \label{eq_ChainHamiltonian}
\end{eqnarray}
where
\begin{eqnarray}
	H_{N+1}&=&\Lambda^{1/2}H_N+\sum_\sigma\bigg[\Lambda^{-1/2}\epsilon_{N+1}c^\dagger_{N+1\sigma}c_{N+1\sigma}+t_N\left(c^\dagger_{N\sigma}c_{N+1\sigma}+c^\dagger_{N+1\sigma}c_{N\sigma}\right)\bigg],\\
	H_0&=&\Lambda^{-1/2}\bigg[\sum_\sigma\epsilon_0c^\dagger_{0\sigma}c_{0\sigma}
	+\frac{1}{\alpha}(H_{\rm{def}}+H_{\rm{hyb}})\bigg],\\
	H_{\rm{hyb}}&=&VF\sum_\sigma\left(c^\dagger_{0\sigma}d_\sigma+{\rm h.c.}\right).
\end{eqnarray}
Here, $c_{n\sigma}$ is an annihilation operator of the conduction electrons on the $n$th site of the chain. 
$\Lambda(>1)$ is a logarithmic discretization parameter and $\alpha=(1+\Lambda^{-1})/2$. 
In the following, $\Lambda$ is set to be $\Lambda=2$. 
$t_N$ is the \lq\lq hopping parameter", and $\epsilon_N$ is the \lq\lq on-site potential", and $F$ is a factor. 
These parameters reflect the DOS of the conduction electrons, and have numerical values of order of unity~\cite{Gonzalez-Buxton1998}. 

By diagonalizing this Hamiltonian iteratively $N$ times and by eliminating higher energy states, we obtain the energy eigenvalues and the eigenstates. 
Here 300 states are retained. 
By using these eigenvalues and eigenstates, thermodynamic quantities are calculated~\cite{Bulla2008}. 
In this paper, the localized moment contributions to the entropy, $S_{\rm{def}}$, and the expectation value of square of the $z$ component of localized moment, $\langle S^2_z\rangle_{\rm{def}}$, are calculated. 

\section{Numerical results}
\begin{figure}
	\centering
	\includegraphics[width=9cm]{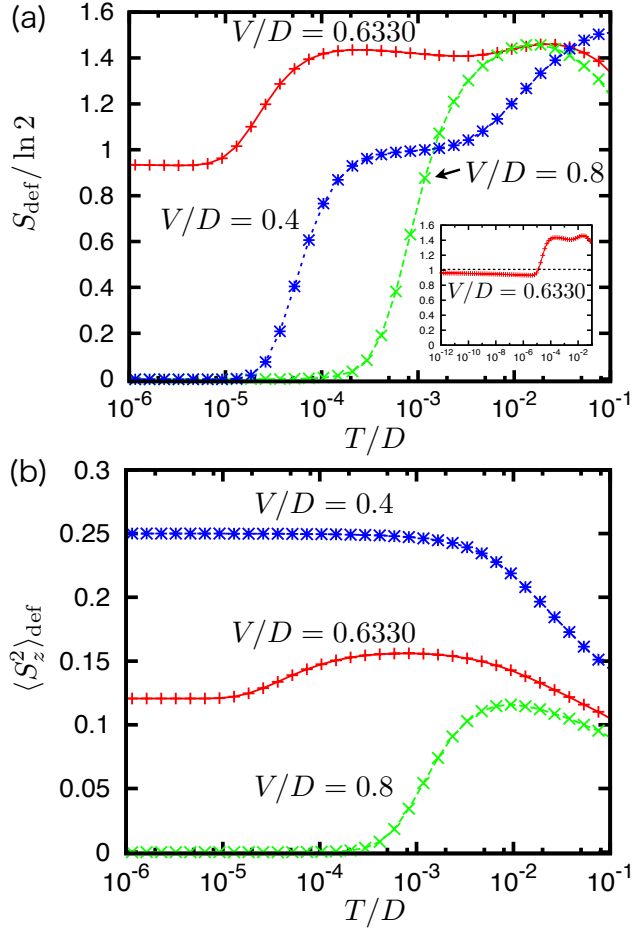}\label{fig_thermodynamic}
	\caption{(a) Temperature dependence of entropy of the electron on the defect $sp^2$ orbital, $S_{\rm{def}}$, under the Zeeman field $h/D=10^{-4}$ at chemical potential $\mu=0$ for the hybridization $V/D=0.4$, $0.6330$, and $0.8$. 
	Here, Boltzmann constant $k_B=1$. 
	Inset shows $S_{\rm def}$ at $V/D=0.6330$ for $10^{-12}<T/D<10^{-1}$. 
	(b) Temperature dependence of expectation value of $S^2_z$, $\langle S^2_z\rangle_{\rm{def}}$, for the same parameters as in Fig. 1(a). }
\end{figure}
In order to clarify the low-temperature states of this model, we calculate the temperature dependence of $S_{\rm def}$ and $\langle S^2_z\rangle_{\rm{def}}$ in the NRG method. 
The parameters $D, \epsilon_{sp^2},$ and $U$ are assumed to be $D\simeq8$ ${\rm eV}$, $\epsilon_{sp^2}\simeq-1$ ${\rm eV}$, and $U\simeq3$ ${\rm eV}$. 
Thus we use the parameters $\epsilon_{sp^2}/D=-0.125$, and $U/D=0.375$, in the following. 
With these parameters, in the case of no Zeeman field, $h/D=0$, it has been shown~\cite{Kanao2012} that the critical hybridization is $V_c/D=0.6311$. 
At $\mu=0$, the entropy of the defect is $S_{\rm{def}}=\ln2$, $\ln3$, and $0$, for $V<V_c, V=V_c$, and $V>V_c$, respectively at low temperatures. 
These three cases correspond to the local moment state, the valence-fluctuation fixed point, and the Kondo-Yosida singlet state, respectively. 
Here, the Boltzmann constant is set to be $k_B=1$.

Figure 1(a) shows $S_{\rm def}$ as a function of temperature for the hybridization $V/D=0.4$, $0.6330$, and $0.8$ under the Zeeman field of $h/D=10^{-4}$ at $\mu=0$. 
For the case with $V/D=0.4$ and $0.8$, $S_{\rm def}\simeq0$ at low temperatures. 
However, their temperature dependences are different: 
At $V/D=0.4$, for $T\simeq10^{-4}D(\simeq h)$, $S_{\rm def}$ changes from $\ln2$ to $\ln1=0$. 
Thus, it is suggested that the ground state is a spin polarized state. 
At $V/D=0.8$, on the other hand, $S_{\rm{def}}$ changes from $\ln3$ to $\ln1$ for $T/D\simeq10^{-3}$. 
This temperature is much larger than the Zeeman field ($h/D=10^{-4}$). 
Thus, the ground state is considered to be the Kondo-Yosida singlet state. 
At a critical point $V/D=0.6330$, we find that $S_{\rm{def}}$ approaches $\ln2$ at the lowest temperature as shown in the inset of Fig. 1(a). 

In order to obtain more information on the electronic states at low temperatures, we calculate $\langle S^2_z\rangle_{\rm{def}}$. 
Figure 1(b) shows $\langle S^2_z\rangle_{\rm{def}}$ as a function of temperature for the same parameters. 
For the case with $V/D=0.4$ and $0.8$, we find $\langle S^2_z\rangle_{\rm{def}}\simeq1/4$ and $0$, respectively at low temperature, which confirms the ground states suggested above: the spin polarized state for $V/D=0.4$, and the Kondo-Yosida singlet state for $V/D=0.8$. 
It is also found that $\langle S^2_z\rangle_{\rm{def}}$ suddenly changes from $1/4$ to $0$ at $V/D=0.6330$ when we change $V$. 
Thus, $V/D=0.6330\simeq{\tilde V}_c/D$ is a critical value which separates the spin polarized state and the Kondo-Yosida singlet state.  
At $V\simeq{\tilde V}_c$, $\langle S^2_z\rangle_{\rm{def}}\simeq1/8$. 

The critical value in the Zeeman field, ${\tilde V}_c$, is larger than the critical value $V_c$ at $h=0$, because the Kondo-Yosida singlet state is suppressed by the Zeeman field. 
We suppose that ${\tilde V}_c$ is determined by the relation $T_K\sim h$, where $T_K$ is a Kondo temperature. 
The values of $S_{\rm{def}}=\ln2$ and $\langle S^2_z\rangle_{\rm{def}}=1/8$ at ${\tilde V}_c$ imply that there are two degenerate states at this point. 
We expect that these two states are up-spin state and empty (unoccupied) state, since $\langle S^2_z\rangle_{\rm{def}}=1/8$ can be interpreted as $[(1/2)^2+0]/2$~\cite{Fritz2004}. 
The detail of this unusual fixed point will be discussed elsewhere. 

We also calculate the finite chemical potential case, $\mu\neq0$, under the Zeeman field. 
$S_{\rm{def}}$ is always zero at low temperatures. 
At low temperatures, $\langle S^2_z\rangle_{\rm{def}}$, shows the change from $\langle S^2_z\rangle_{\rm{def}}=1/4$ to $0$, as $V$ increases. 
We do not find any unusual fixed point in this case. 

\section{Conclusion}
In summary, the effect of the Zeeman field on the defect-induced Kondo effect in graphene has been investigated. 
The effective model with the Zeeman field has been derived and has been analyzed on the basis of numerical renormalization group method. 
It has been found that, at $\mu=0$, there is an unusual fixed point with the entropy of $k_B\ln2$ and the expectation value of $S^2_z$ of $1/8$ at the border of the spin polarized state and the Kondo-Yosida singlet state. 

\ack
T. K. is supported by the Global COE program \lq\lq the Physical Sciences Frontier" of the Ministry of Education, Culture, Sports, Science and Technology, Japan.

\end{document}